\documentclass{amsart}

\usepackage[utf8]{inputenc}
\usepackage[usenames,dvipsnames]{color}
\usepackage{listings}
\usepackage{amsmath}
\usepackage{amssymb}
\usepackage{amsthm}
\usepackage{tikz}

\usepackage[pdfborder={0 0 0 0},colorlinks=true]{hyperref}

\DeclareSymbolFont{rsfscript}{OMS}{rsfs}{m}{n}
\DeclareSymbolFontAlphabet{\mathrsfs}{rsfscript}

\DeclareFontFamily{OMS}{rsfs}{\skewchar\font'177}
\DeclareFontShape{OMS}{rsfs}{m}{n}{%
      <5> rsfs5
      <6> <7> rsfs7
      <8> <9> <10> rsfs10
      <10.95> <12> <14.4> <17.28> <20.74> <24.88> rsfs10
      }{}

\def\calF{\mathrsfs{F}}
\def\calG{\mathrsfs{G}}

\def\calJ{\mathrsfs{J}}

\def\calO{\mathrsfs{O}}
\def\calP{\mathrsfs{P}}
\def\calQ{\mathrsfs{Q}}
\def\calR{\mathrsfs{R}}

\def\calV{\mathrsfs{V}}

\DeclareMathOperator{\lt}{lt}

\DeclareMathOperator{\Fin}{Fin}
\DeclareMathOperator{\Ord}{Ord}
\DeclareMathOperator{\Vect}{Vect}

\theoremstyle{plain}

\newtheorem {proposition}{Proposition}

\theoremstyle{definition}

\newtheorem {definition}{Definition}
\newtheorem {remark}{Remark}

\title{Implementing Gr\"obner bases for Operads}
\author{Vladimir Dotsenko}
\address{Dublin Institute for Advanced Studies, 10 Burlington Road, Dublin 4, Ireland and School of Mathematics, Trinity College, Dublin 2, Ireland}
\email{vdots@maths.tcd.ie}
\author{Mikael Vejdemo Johansson}
\address{Department of Mathematics, Stanford CA 94305, USA}
\email{mik@stanford.edu}

\begin{document}

\lstset{language=Haskell}

\begin{abstract}
We present an implementation of the algorithm for computing
Gr\"obner bases for operads due to the first author and A.~Khoroshkin. We
discuss the actual algorithms, the choices made for the implementation
platform and the data representation, and strengths
and weaknesses of our approach.
\end{abstract}

\maketitle

\section{Introduction}
\label{sec:introduction}

\subsection{Summary of results}
\label{sec:sum-of-results}

In an upcoming paper~\cite{dotsenko_grobner_2008}, the first author
and Anton Khoroshkin define the concept of a Gr\"obner basis for
finitely presented operads. In that paper, they prove the diamond
lemma, and demonstrate that for an operad, having a quadratic
Gr\"obner basis is equivalent to the existence of a
Poincar\'e--Birkhoff--Witt basis. As demonstrated by Eric
Hoffbeck~\cite{hoffbeck_poincar-birkhoff-witt_2007}, an operad with a PBW basis is Koszul.  Hence, an
implementation of the Gr\"obner bases algorithm yields,
in addition to a framework for exploration of operads by means
of explicit calculation, a computer-aided tool for proving
Koszulness.

In this paper, we present an implementation of the Gr\"obner basis algorithm in
the Haskell programming language~\cite{haskell98}. Being designed with categorical 
terms, Haskell provides a powerful framework for algorithms like that. What we end up
with is a computer sofware package which allows to compute the Gr\"obner basis for
a finitely presented operad, as well as bases and dimensions for components of such
an operad. 

One of the main goals of this paper is to help mathematicians who want to get familiar with this
software package and use it for their needs, including changing some algorithms or adding
more functionality.\footnote{The first author is a living example proving that it is possible; having been
introduced to Haskell by the second author in the process of working on this package, he now has
enough confidence to not only use the package, but to add new functions as well.} Consequently, this is more
of an invitation to experiment with this software than a report on what it is possible to compute. Let
us comment briefly on the state of the art regarding computations. While working on the package, we have 
implemented several well known operads to test the performance. In the case when an operad is PBW, our package 
captures that right away. This 
already is a very important achievement: having implemented many different admissible orderings, one can 
check very fast whether or not an operad is PBW for at least one of them, thus proving the Koszulness in many
cases. Note that the PBW property depends a lot not only on the choice of an admissible ordering, but also on
the choice of ordering of generators of our operad; for example, for the operad of pre-Lie algebras, depending on
the ordering, a Gr\"obner basis can vary from quadratic to seemingly infinite. On the other hand, for operads that do not 
have a quadratic Gr\"obner basis, we encountered subtle performance issues in many cases. For operads having a 
relatively small finite Gr\"obner basis, like the fake commutative operad $\mathop{\mathrm{AntiCom}}$ \cite{dotsenko_grobner_2008},
the computation easily yields the correct result, while for many other cases, like the pre-Lie operad for a ``wrong'' ordering, 
computations with arity~$6$ and further take enormously long. 

The actual implementation is distributed through the HackageDB repository for Haskell software projects 
at \url{http://hackage.haskell.org/package/Operads}~--- software distributed through this repository are 
available through the automated installation tool \verb+cabal-install+.

The current documentation files are kept online at \url{http://math.stanford.edu/~mik/operads/}. 

\subsection{Outline of the paper}
\label{sec:outline}

The paper is organized as follows. In Section~\ref{sec:overview}, we recall relevant background
information related to operads and Gr\"obner bases, on one hand, and to types and functions in Haskell, on the other hand. 
In Section~\ref{sec:intern-repr}, we discuss the way we chose to represent our data in Haskell. In Section~\ref{sec:algorithms},
we present algorithms used in our implementation. Finally, in the appendix, we list Haskell constructions used throughout the paper.

\subsection{Acknowledgements}
\label{sec:acknowledgements}

We wish to express our deep gratitude to Eric Hoffbeck and Henrik
Strohmayer for both significant assistance in the construction of the
software code, and analysis of the techniques we are using. Some of
the hairier points of Haskell evaluation has been rendered clear by
the helpful assistance of the many members of the \verb+#haskell+ IRC
channel on the Freenode IRC network.

The first author was supported by an IRCSET research fellowship. The second author was supported by the Office of Naval Research, through grant 
N00014-08-1-0931.

We are grateful to Jean--Louis Loday and Bruno Vallette who organized the ``Operads 2009'' meeting in CIRM Luminy, where the work on this project was started. The second author wishes to thank Dublin Institute for Advanced Studies which hosted him as a visitor during the last stage of working on this paper.

\section{Overview}
\label{sec:overview}

For exhaustive information on symmetric operads, we refer the reader to monographs~ \cite{loday-vallette_algebraic_operad} and~\cite{markl_operads_2002}. Here,
we mainly concentrate on shuffle operads, and their relationship with symmetric operads, and definitions in the symmetric case are chosen in the way that
best suits this approach.

\subsection{Operads} \label{sec:operads} 
We denote by~$\Ord$ the category of nonempty finite ordered sets (with order-preserving bijections as morphisms), 
and by $\Fin$~--- the category of nonempty finite sets (with bijections as morphisms). Also, we denote by $\Vect$ the category of vector spaces (with 
linear operators as morphisms; unlike the first two cases, we do not require a map to be invertible).

\begin{definition}
\begin{enumerate}
 \item A \emph{(nonsymmetric) collection} is a contravariant functor from the category~$\Ord$ to the category~$\Vect$.
 \item A \emph{symmetric collection} (or an \emph{$\mathbb{S}$-module}) is a contravariant functor from the category~$\Fin$ to the category~$\Vect$.
\end{enumerate}
For either type of collections, we can consider the category whose objects are collections of this type (and morphisms are morphisms of the corresponding 
functors). The natural forgetful functor ${}^{f}\colon\Ord\to\Fin$, $I\mapsto I^f$ leads to a forgetful functor ${}^f$ from the category of symmetric 
collections to the category of nonsymmetric ones, $\calP^f(I):=\calP(I^f)$. For simplicity, we let $\calF(k):=\calF([k])$.
\end{definition}

We use the convention $[k]=\{1,2,\dots,k\}$ in this paper. 

\begin{definition}
\begin{itemize}
\item Let $\calP$ and $\calQ$ be two nonsymmetric collections. Define their \emph{shuffle composition} $\calP\circ_{sh}\calQ$ by the formula
 $$
(\calP\circ_{sh}\calQ)(I):=\bigoplus_{k}\calP(k)\otimes\left(\bigoplus_{f\colon I\twoheadrightarrow[k]}\calQ(f^{-1}(1))\otimes\ldots\otimes\calQ(f^{-1}(k))\right),
 $$
where the sum is taken over all shuffling surjections~$f$, that is surjections for which~$\min f^{-1}(i)<\min f^{-1}(j)$ whenever~$i<j$.
\item Let $\calP$ and $\calQ$ be two symmetric collections. Define their \emph{(symmetric) composition} $\calP\circ\calQ$ by the formula
 $$
(\calP\circ\calQ)(I):=\bigoplus_{k}\calP(k)\otimes_{S_k}\left(\bigoplus_{f\colon 
I\twoheadrightarrow[k]}\calQ(f^{-1}(1))\otimes\ldots\otimes\calQ(f^{-1}(k))\right),
 $$
where the sum is taken over all surjections~$f$.
\end{itemize}
\end{definition}

Each of these compositions gives a structure of a monoidal category on the category of the corresponding collections. The same definitions can be given if 
we replace $\Vect$ by another symmetric monoidal category. For our purposes, an important replacement for $\Vect$ will be the category of finite sets (with 
arbitrary mappings as morphisms).

\begin{definition}
\begin{enumerate}
 \item A \emph{shuffle operad} is a monoid in the category of nonsymmetric collections with the monoidal structure given by the shuffle composition. 
 \item A \emph{symmetric operad} is a monoid in the category of symmetric collections with the monoidal structure given by the (symmetric) composition.
\end{enumerate}
\end{definition}

\begin{definition}
A \emph{shuffle permutation} of the type $(k_1,\ldots,k_n)$ is a permutation in the symmetric group $S_{k_1+\ldots+k_n}$
 which preserves the order of the first $k_1$ elements, the second $k_2$ elements,\ldots, the last $k_n$ elements, and satisfies  
 $$
\sigma(1)<\sigma(k_1+1)<\sigma(k_1+k_2+1)<\ldots<\sigma(k_1+\ldots+k_{n-1}+1).
 $$
\end{definition}

\begin{proposition}
The number of shuffle permutations of the type $(k_1,\ldots,k_n)$ is equal to 
 $$
\frac{k_1k_2\cdot\ldots\cdot k_n}{(k_1+k_2+\ldots+k_n)(k_2+\ldots+k_n)\cdot\ldots\cdot k_n}\binom{k_1+k_2+\ldots+k_n}{k_1,k_2,\ldots,k_n}. 
 $$ 
\end{proposition}
 
When implementing shuffle permutations, one can use the following simple idea: In a shuffle permutation, the number whose image is $k_1+\ldots+k_n$ should 
clearly be the maximal one in its block. Moreover, if this block is of size~$1$, it should be the last one to comply with the ordering condition on the 
first elements of blocks. This implies an obvious recursive algorithm to generate a list of shuffle permutations with given sizes of blocks: put the 
maximal image in the end of each allowed block, and for each such choice list all shuffle permutations where the corresponding block contains one element 
less than prescribed.

\begin{definition}
\begin{enumerate}
 \item Let $\calO$ be a shuffle operad, $\beta\in\calO(n)$, $\alpha_1\in\calO(k_1)$, \ldots, $\alpha_n\in\calO(k_n)$. Assume that $\sigma\in 
S_{k_1+\ldots+k_n}$ is a shuffle permutation of the type $(k_1,\ldots,k_n)$. Denote by $B_s$, $s=1,\ldots,n$, the $s^{\text{th}}$ block of $[k_1+\ldots+k_n]$ 
(on which $\sigma$ is monotonous). Then we define
  $$
 \beta(\alpha_1,\ldots,\alpha_n)_\sigma=\circ(\beta\otimes\sigma(\alpha_1)\otimes\ldots\otimes\sigma(\alpha_n))\in\calO(k_1+\ldots+k_n),
  $$
where $\sigma(\alpha_s)$ is the image of $\alpha_s$ under the isomorphism between $\calO(k_s)$ and $\calO(\sigma(B_s))$, and 
$\circ\colon\calO\circ_{sh}\calO\to\calO$ is the monoid product map.
 \item Let $\calO$ be a symmetric operad, $\beta\in\calO(n)$, $\alpha_1\in\calO(k_1)$, \ldots, $\alpha_n\in\calO(k_n)$. Let $\sigma\in S_{k_1+\ldots+k_n}$ be an 
arbitrary permutation. Denote by $B_s$, $s=1,\ldots,n$, the $s^{\text{th}}$ block of $[k_1+\ldots+k_n]$ (of size $k_s$). Then we define
  $$
 \beta(\alpha_1,\ldots,\alpha_n)_\sigma=\circ(\beta\otimes\sigma(\alpha_1)\otimes\ldots\otimes\sigma(\alpha_n))\in\calO(k_1+\ldots+k_n),
  $$
where $\sigma(\alpha_s)$ is the image of $\alpha_s$ under the isomorphism between $\calO(k_s)$ and $\calO(\sigma(B_s))$, and 
$\circ\colon\calO\circ\calO\to\calO$ is the monoid product map.
\end{enumerate} 
\end{definition}

It turns out that the forgetful functor is a monoidal functor between the category of symmetric operads and the category of shuffle operads. Consequently, 
it turns out that to study various questions of linear algebra for symmetric operads, it is sufficient to forget the full symmetric structure because the 
shuffle structure already captures everything. See~\cite{dotsenko_grobner_2008} for more details.

\subsection{Trees}
\label{sec:trees}

Assume that we are given a collection of disjoint finite sets $M(n),
n\ge1$. A \emph{(rooted) tree} is a non-empty directed graph $T$ of
topological genus 0 
for which each vertex has at least one incoming edge and exactly one outgoing edge. We allow for some edges of a tree to be bounded by a vertex 
at one end only. Such edges are called \emph{external}. Each tree has 
one outgoing external edge, the \emph{output} or the \emph{root}, and several ingoing external edges, called \emph{leaves}. All vertices of the 
tree should be decorated by elements of sets from the collection $M$; for a vertex with $n$ inputs, the element used for the decoration should belong 
to~$M(n)$. Such a tree will be referred to as a nonsymmetric tree monomial. Usually, we consider such trees with some additional structure: for a tree 
with~$n$ leaves, we require the leaves to be labelled by~$[n]$. For each vertex $v$ of a tree, the edges going in and out of $v$ will be referred to as 
inputs and outputs at~$v$. A tree with a single vertex is called a \emph{corolla}. There is also a tree with a single input and no vertices called 
the \emph{degenerate} tree. Trees are originally considered as abstract graphs but to work with them we would need some particular representatives that we 
now are going to describe.

For a tree with labelled leaves, its canonical planar representative is defined as follows. In general, an embedding of a (rooted) tree in the plane is 
determined by an ordering of inputs for each vertex (the leftmost one being the smallest, the rightmost~--- the largest). To compare two inputs of a vertex~$v$, we find the minimal leaves that one can reach from~$v$ via the 
corresponding input. The input for which the minimal leaf is smaller is considered to be less than the other one. Planar representatives of decorated trees 
will be referred to as \emph{tree monomials}. The collection of all tree monomials whose vertices are labelled by the collection~$M$ 
is denoted by~$\calF_M$.

Compositions of trees are defined as follows. Given a tree $\beta$ with $n$ leaves, trees $\alpha_1$, \ldots, $\alpha_n$ with $k_1$, \ldots, $k_n$ leaves respectively, we 
define the composition $\beta(\alpha_1,\alpha_2,\ldots,\alpha_n)$ as the tree obtained by grafting the tree $\alpha_1$ to the first leaf of $\beta$, the tree $\alpha_2$ with leaf labels 
shifted by $k_1$ to the second leaf of $\beta$, \ldots, the tree $\alpha_n$ with leaf labels shifted by $k_1+\ldots+k_{n-1}$ to the last leaf of $\beta$. If, in 
addition, $\sigma$ is a shuffle permutation of $[k_1+\ldots+k_n]$ of the type $(k_1,\ldots,k_n)$, we can define the shuffle composition 
$\beta(\alpha_1,\alpha_2,\ldots,\alpha_n)_\sigma$; to compute it, we first compute the nonsymmetric composition of our trees and then apply~$\sigma$ to the leaf labels of the 
resulting tree.

\begin{remark}
Let us emphasize two practicalities. First of all, the data type for trees chosen for an implementation should be easily adjustable for performing 
compositions. Second, it is important here that we apply the permutation, not its inverse; usually, action of permutations on functions and mappings uses 
inverses, however, since operads are contravariant functors, we do not need that. One should be careful, and remember it when implementing the action of 
permutations.
\end{remark}

\begin{proposition}[\cite{dotsenko_grobner_2008}]
\begin{enumerate}
 \item The collection of trees $\calF_M$ is closed under shuffle compositions.
 \item Every tree in $\calF_M$ can be obtained from corollas by iterated shuffle compositions.
\end{enumerate}
\end{proposition}

The collection $\calF_M$ is the free shuffle operad generated by the nonsymmetric collection~$M$ in the category of finite sets; its linear span is the 
free shuffle operad generated by the linear span of~$M$ in the category of vector spaces.

Here and below by a divisor of a nonsymmetric tree monomial~$T$ we mean a nonsymmetric tree monomial~$T'$ whose underlying tree is embedded into the 
underlying tree of~$T$ in such a way that the labellings of vertices are the same.

For a tree monomial~$\alpha$ with the underlying nonsymmetric monomial~$T$ and a divisor~$T'$ of~$T$, let us define a tree monomial~$\alpha'$ that 
corresponds to~$T'$. Its vertices are already decorated, so we just need to take care of the leaf labelling. For each leaf $l$ of $T'$, let us 
consider the smallest leaf of $T$ that can be reached from~$l$. We then number the leaves according to these ``smallest descendants'': the leaf with the 
smallest possible descendant gets the label~$1$, the second smallest~--- the label~$2$ etc.

\begin{definition}
For two tree monomials $\alpha, \beta\in \calF_\calV$, we say that \emph{$\alpha$ is divisible by $\beta$}, if there exists a nonsymmetric divisor of 
$\alpha$ for which the corresponding tree monomial $\alpha'$ is equal to~$\beta$.
\end{definition}

\begin{remark}
Checking divisibility is important for the Gr\"obner bases algorithm. Thus, one has to put effort in finding an efficient implementation. Our approach is 
recursive; it is very much motivated by the choice of the platform. It would be very interesting to find a reasonably fast non-recursive algorithm for 
that, in the spirit of algorithms of Knuth--Morris--Pratt~\cite{knuth_morris_pratt} and Boyer--Moore~\cite{boyer_moore} for string divisibility.
\end{remark}

\begin{proposition}[\cite{dotsenko_grobner_2008}]
A tree monomial $\alpha$ is divisible by $\beta$ if and only if $\alpha$ can be obtained from $\beta$ by iterated shuffle compositions with corollas. 
\end{proposition}

Assume that $\alpha$ is divisible by $\beta$. Take some sequence of shuffle compositions with corollas that produces $\alpha$ from $\beta$. This sequence 
can be applied to any tree monomial with the same number of arguments as~$\beta$; abusing the notation a little bit, we denote that operation on tree 
monomials by $m_{\alpha,\beta}$. This operation is actually well
defined (that is, depends only on~$\alpha$ and the specific subtree corresponding to the 
divisor~$\beta$, but not on any particular sequence of compositions that creates $\alpha$ from~$\beta$).

\begin{remark}
\label{rem:black-hole}
Being able to compute the operations $m_{\alpha,\beta}$ in the most efficient way is crucial for the Gr\"obner bases algorithm. This means that much 
thought should be put in the data representation philosophy; keeping the divisibility information in a logical way helps to be efficient. We chose the 
approach where a divisor is replaced by a ``black hole'' (or, in other words, by a new corolla); this way the operation $m_{\alpha,\beta}$ corresponds to 
the simple insertion of a tree in the hole. We explain it in more details in Section \ref{sec:trees-with-holes-2}.
\end{remark}

\begin{definition}
A tree monomial $\gamma$ is called a \emph{common multiple} of two tree monomials $\alpha$ and $\beta$, if it is divisible by both $\alpha$ and $\beta$. 
Tree monomials $\alpha$ and $\beta$ are said to have a \emph{small common multiple}, if they have a common multiple for which the number of vertices of the underlying tree is less than the total number of vertices for $\alpha$ and $\beta$, and the embeddings of $\alpha$ and $\beta$ in this 
common multiple cover all its vertices.
\end{definition}

\begin{remark}
Computation of small common multiples is one of the most frequently used operations in the Gr\"obner basis algorithm. Our approach (described in detail 
below) shares certain similarities with the algorithm that lists all shuffle permutations of the given type, even though is a little bit more 
sophisticated.
\end{remark}

\subsection{Gr\"obner bases}
\label{sec:grobner-bases}

In this section, we assume that we are working with shuffle operads over $\Vect$, in particular, we assume that the set of relations~$\calR$ consists of linear 
combinations of tree monomials. For the case of symmetric operads, pre-processing of the input data is required: first, the symmetric group action should 
be used to express operadic monomials in terms of tree monomials, second, the relations should generate the corresponding ideal as a shuffle ideal. For 
that, it is necessary that they form a symmetric subcollection, that is, are closed under the symmetric groups action. For example, if we add to the set of 
relations the orbit of each relation, this condition is automatically satisfied.

In this section, we work with the free operad $\calF_\calV$, where the nonsymmetric collection~$\calV$ is endowed with a basis, a collection of sets~$M$.

An ordering of tree monomials of $\calF_\calV$ is said to be \emph{admissible}, if it is compatible with the operadic structure, that is, replacing the operations in any shuffle composition with larger operations of the same arities increases the result of the composition. In Section~\ref{sec:monomial-orderings}, we shall describe some admissible orderings. All results below are valid for every admissible ordering of tree monomials.

\begin{definition}
For an element~$\lambda$ of the free operad~$\calF_\calV$, the tree monomial~$\alpha$ is said to be its \emph{leading term}, if it is the maximal monomial 
that occurs in the expansion of $\lambda$ with a nonzero coefficient (notation: $\lt(\lambda)=\alpha$). This nonzero coefficient (the leading coefficient 
of $\lambda$) is denoted by $c_\lambda$.
\end{definition}

\begin{remark}
Whereas in the previous section we only worked with trees, from now on we use linear combinations of trees. Thus, it is important to implement working with 
tree polynomials, that is, linear combinations of tree monomials. The main requirement is that obtaining the leading term and the leading coefficient, 
being the most frequently used operations on polynomials, should be easy.
\end{remark}

\begin{definition}
Assume that $f$ and $g$ are two elements of $\calF_\calV$ for which the leading term of $f$ is divisible by the leading term of $g$. The element
 $$
r_g(f):=f-\frac{c_f}{c_g}m_{\lt(f),\lt(g)}(g),
 $$
is called the \emph{reduction of $f$ modulo~$g$}. 
\end{definition}

\begin{definition}
Assume that $f$ and $g$ are two elements of $\calF_\calV$ whose leading terms have a small common multiple $\gamma$. We have
 $$
m_{\gamma,\lt(f)}(\lt(f))=\gamma=m_{\gamma,\lt(g)}(\lt(g)).
 $$
The element
 $$
s_\gamma(f,g):=m_{\gamma,\lt(f)}(f)-\frac{c_f}{c_g} m_{\gamma,\lt(g)}(g),
 $$
is called the $S$-polynomial of $f$ and $g$ (corresponding to the common multiple $\gamma$; note that there can be several different small common multiples). 
\end{definition}

$S$-polynomials, as defined here, include the reductions as a particular case. It turns out to be convenient, but we shall need reductions on their own to deal with Gr\"obner bases.

\begin{definition}
Let $\calJ$ be an ideal of the free operad. $\calG$ is called a \emph{Gr\"obner basis} of~$\calJ$, if for every $f\in\calJ$ the leading term of $f$ is divisible by the leading term of some element of~$\calG$.
\end{definition}

The main fact about Gr\"obner bases that makes them so useful is that the tree monomials that are not divisible by leading terms of a Gr\"obner basis form a basis for the quotient of the free operad modulo~$\calJ$. Thus, knowing a Gr\"obner basis for defining relations of an operad allows to obtain important information about this operad.

Recall the Buchberger's algorithm for operads~\cite{dotsenko_grobner_2008}. Its input is a set~$\calR$ of relations between elements of the free shuffle 
operad~$\calF_\calV$. It repeatedly applies the following step: 

\begin{quote}
\textbf{Step of the Buchberger's algorithm:} Compute all pairwise $S$-polynomials of elements of~$\calR$. Reduce all these elements modulo~$\calR$ until 
they cannot be reduced further. Extend~$\calR$ by joining these reductions to it. If there are no new elements joined, terminate. (If there are new 
elements, the step is repeated for the newly obtained set~$\calR$.)
\end{quote}

\subsection{Haskell}
\label{sec:haskell}

Haskell~\cite{haskell98} is a purely functional programming language
with a powerful type system. Programming in Haskell has a declarative
feel to it, in the sense that functions are defined by declaring
equations for function evaluation~--- the equations are then used by
the compiler with the first matching equation in the source code used
for every application in the source code.

We have built the implementation we are discussing in Haskell, and
will use occasional source code excerpts for illustration through the
paper. See the appendix for the list of all the
Haskell-specific functions in use in our code examples.

\subsubsection{Types}
\label{sec:types}

Haskell depends on a strong adherence to its type system. Hence, any
entity in the language possesses a type. There are types that are
complete in themselves~--- such as \lstinline!Bool!, \lstinline!Int!,
and types that are assembled from component types~--- such as the type
\lstinline![a]! for lists containing elements of the type
\lstinline!a!. Functions, too, are first class citizens of the
language, with a function taking input of type \lstinline!a! and
returning output of type \lstinline!b! having type \lstinline!a -> b!.

Real power in the Haskell treatment of data types appears with the
freedom to declare your own type. The most complete way to do this is
with the \lstinline!data! declaration. This allows, easily, for both
record and union types; where a record contains one value of each of
the specified values, and a union contains on value out of the
specified values.

As an example, similar to the datatype for trees that we will discuss
at length in Section \ref{sec:decorated-trees}, a rooted tree is
either a leaf node, or a root node with a list of subtrees. The arity
of the root node will be precisely the length of the list of
subtrees. And a node of arity 0 could be considered a leaf.

Thus, we may define a tree data type using the declaration
\begin{lstlisting}
  data Tree = Leaf | Node [Tree]
\end{lstlisting}

\begin{figure}[here]
\centering
\begin{tikzpicture}
  \node [shape=circle, minimum size=5mm, draw] (n0) at (0,2)     {} ;
  \node [shape=circle, minimum size=5mm, draw] (n1) at (-1,1)    {} ;
  \node [shape=circle, minimum size=5mm, draw] (n2) at (1,1)     {} ;
  \node  (l1) at (-1.5,0)  {} ;
  \node  (l2) at (-0.5,0)  {} ;
  \node  (l3) at (0.5,0)   {} ;
  \node  (l4) at (1.0,0)   {} ;
  \node  (l5) at (1.5,0)   {} ;
  \draw [->] (n0) to (n1) ;
  \draw [->] (n0) to (n2) ;
  \draw [->] (n1) to (l1) ;
  \draw [->] (n1) to (l2) ;
  \draw [->] (n2) to (l3) ;
  \draw [->] (n2) to (l4) ;
  \draw [->] (n2) to (l5) ;
\end{tikzpicture}
\caption{A tree shape} 
\label{fig:treeshape1}
\end{figure}
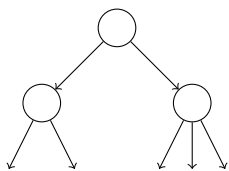

Here, \lstinline!Tree! is the resulting data type, and
\lstinline!Leaf! and \lstinline!Node! are constructors for elements of
the data type. A typical tree might look like:

\begin{lstlisting}
  Node [Node [Leaf, Leaf], Node [Leaf, Leaf, Leaf]]
\end{lstlisting}

The corresponding tree shape is shown in Figure~\ref{fig:treeshape1}.

This data type has been defined completely using the union type: a
tree is either a leaf, or a node carrying a list of subtrees. We can
extend the type using the record type construction into something that
can carry labels both on leaves and vertices, making it usable
to represent the decorated trees we use for representing operads.

Thus, we can introduce two \emph{type variables}, to make the resulting
tree type versatile, and define a tree type that takes labels of any
type for the nodes, and labels of any type~--- independent of the node
type~--- for the leaves by:

\begin{lstlisting}
data Tree a b = Leaf a | Node b [Tree]
\end{lstlisting}

Here, an element of the type \lstinline!Tree! is either a leaf,
equipped with a value of type \lstinline!a!, or a node, equipped with
a value of type \lstinline!b! as well as a list of subtrees.

\begin{figure}[here]
\centering
\begin{tikzpicture}
  \node [shape=circle, minimum size=5mm, draw] (n0) at (0,2)     {2} ;
  \node [shape=circle, minimum size=5mm, draw] (n1) at (-1,1)    {1} ;
  \node [shape=circle, minimum size=5mm, draw] (n2) at (1,1)     {3} ;
  \node  (l1) at (-1.5,0)  {'a'} ;
  \node  (l2) at (-0.5,0)  {'b'} ;
  \node  (l3) at (0.5,0)   {'c'} ;
  \node  (l4) at (1.0,0)   {'d'} ;
  \node  (l5) at (1.5,0)   {'e'} ;
  \draw [->] (n0) to (n1) ;
  \draw [->] (n0) to (n2) ;
  \draw [->] (n1) to (l1) ;
  \draw [->] (n1) to (l2) ;
  \draw [->] (n2) to (l3) ;
  \draw [->] (n2) to (l4) ;
  \draw [->] (n2) to (l5) ;
\end{tikzpicture}
\caption{A decorated tree} 
\label{fig:treeshape2}
\end{figure}
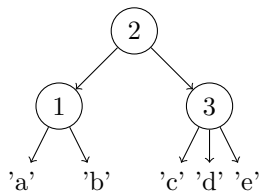

Hence, an example of type \lstinline!Tree Char Int! would be:

\begin{lstlisting}
Node 2 [Node 1 [Leaf 'a', Leaf 'b'], 
        Node 3 [Leaf 'c', Leaf 'd', Leaf 'e']]
\end{lstlisting}

The corresponding decorated tree is shown in Figure~\ref{fig:treeshape2}.

\subsubsection{Functions}
\label{sec:functions}

The second important part of understanding the way Haskell works is
the functions. A function is defined by its type, and by what it makes
to the input arguments it takes. Haskell views a function with several
input parameters as a function taking one value and returning a
function expecting one less parameter. Thus \lstinline!(+)! is a
function that when applied to the value \lstinline!2! will return a
function \lstinline!(2+)! that in turn increases its parameter by 2.

A function specification in Haskell has two components. First off is
the (optional) type declaration. An example, taken from our source code: 

\begin{lstlisting}
operadicBuchberger :: 
        (Ord a, Show a, TreeOrdering t, Fractional n) => 
        [OperadElement a n t] -> [OperadElement a n t]
\end{lstlisting}

This type declaration alone tells us a number of things about the
function, and the parameters it takes and returns. First comes the
name of the function: \lstinline!operadicBuchberger!. It is the top
level function to run the Buchberger algorithm on a set of operadic
relations. Next is the \lstinline!::!~--- signifying that a type
declaration follows.

Following the name and the \lstinline!::!, we give an (optional) list of assumptions
on the type variables involved in constructing all types of the type
declaration: we need \lstinline!a! to be a type that can be sorted and
printed, we need \lstinline!t! to be a \lstinline!TreeOrdering!,
i.e. an implementation of the monomial ordering algorithms we
use. Finally, we expect \lstinline!n! to be a numeric type
implementing a field.

Following the expectations follows the symbol \lstinline!=>!,
signifying the start of the actual type declaration. And we read off
that the function has type \\
\lstinline![OperadElement a n t] -> [OperadElement a n t]!, or in
other words that \\
\lstinline!operadicBuchberger! takes a list of operad
elements with a certain type of vertex labels (signifying the
operations in the free operad), a certain type of coefficients and a
certain monomial ordering.

After the type declaration, a sequence of equational declaration
follow, containing the bulk of the function definition. As a function
is called, these equations are processed in the order they are stated
until one is found such that the left hand side of the equation
matches the parameters submitted to the call. Once a match is found,
the code on the right hand side is executed and the result is returned
as the value of the function call.

Again, an example may be in order. We can write a function for the
\lstinline!Tree! type we described above that allow us to recognize
leaves:

\begin{lstlisting}
isLeaf :: Tree a b -> Bool
isLeaf (Leaf leaflabel) = True
isLeaf (Node nodelabel subtrees) = False
\end{lstlisting}

\section{Internal representations}
\label{sec:intern-repr}

Data representations and algorithms go hand in hand. A good algorithm
will suggest a data representation that makes the steps of that
algorithm easier; a good data representation will make the algorithms
handling the data obvious and efficient from the storage methods. We
have tried to find representations for the data types required for the
Gr\"obner basis algorithms that will make the subtasks we have
identified easy to implement efficiently.

We discuss representation for decorated trees in Section
\ref{sec:decorated-trees}. The elements of a free operad in the
category $\Vect$ are formal linear combinations of decorated trees,
and the representation of these is discussed in Section
\ref{sec:operad-elements}. Next up is the special type trickery needed
to represent the \emph{black hole} trees first introduced in Remark
\ref{rem:black-hole} on page \pageref{rem:black-hole}. We introduce
two coproduct types useable for tagging vertices of trees while
preserving some or all of the vertex tags in Section
\ref{sec:an-aside-data}. This way, we can designate a corolla a black
hole, or an embedding point in a small common multiple. We discuss the
small common multiple structures in Section
\ref{sec:trees-with-marked} and the black hole tagging in Section
\ref{sec:trees-with-holes-2}. Finally, we discuss the representation
we use for permutations in Section \ref{sec:permutations}.

\subsection{Decorated trees}
\label{sec:decorated-trees}

We recall that the free operad is built with trees decorated at the
corollas with elements of the generating graded set and with leaves
decorated with an ordered set.

While we expect the trees representing the basis of a free operad to
have integer leaf labels, some of the lower level tasks are easier if
we can also represent trees with other types of leaf labels. Hence, we
will define one underlying tree type, \lstinline!PreDecoratedTree!,
and derive another tree type, \lstinline!DecoratedTree! representing
the basis elements of free operads.

Hence, we will build our software with the decorated tree as our
fundamental building block. We represent trees using a data type that
encodes corollas and leaves as different, allowing each to carry a
label. This guides us to the data type declaration

\begin{lstlisting}
data (Ord a, Show a) => 
    PreDecoratedTree a b = 
        DTLeaf !b | 
        DTVertex {
           vertexType :: !a, 
           subTrees :: ![PreDecoratedTree a b]}
        deriving (Eq, Ord, Read, Show)
\end{lstlisting}

This is essentially the same as the tree type we discussed at the end
of \ref{sec:types}. It is decorated with more expectations, and some
Haskell idioms to automatically generate functionality. Hence, the
\lstinline!deriving! clause will make the tree type automatically
allow equality checks, sorting and methods to serialize and
deserialize the data into strings. The sorting induced by the
\lstinline!deriving! clause, however, is not in general a monomial
ordering, and we introduce further types in the code to introduce
admissible monomial ordering.

Furthermore, the \lstinline!vertexType! and \lstinline!subTrees!
clauses automatically generate function that allow us to extract the
node label and the list of subtrees from a corolla.

Since we occasionally, but not very often, will feel a need to
decorate the leaves with something different from integers, we define
our tree type as a different type from the type we use with the users
in the end. We define a type synonym \lstinline!DecoratedTree a! that
stands for \lstinline!PreDecoratedTree a Int!, so that the user should
only ever have to see \lstinline!DecoratedTree! occuring.

We also define a number of utility functions for operations on
these trees: methods to apply a function to each node label, and to
apply a function to each leaf label, as well as functions to easily
construct corollas and leaves, and to easily determine the arity of a
corolla and the total number of leaves of a tree.

One pattern that reoccurs a lot here is the basic tree recursion
shape. We give, here, as an example, the code to apply a function to
each vertex label:

\begin{lstlisting}
vertexMap :: 
    (Ord a, Show a, Ord b, Show b) => 
    (a -> b) -> PreDecoratedTree a c -> PreDecoratedTree b c
vertexMap f (DTLeaf i) = DTLeaf i
vertexMap f (DTVertex t ts) = 
    DTVertex (f t) (map (vertexMap f) ts) 
\end{lstlisting}

There is some boiler plate~--- the type variables \lstinline!a! and
\lstinline!b! have to match the assumptions needed to build a
tree. But then the code just states that applying $f$ to the node
labels if you encounter a leaf just returns the leaf unchanged. If,
however, you encounter a vertex, you apply the function to the
label and construct a new vertex with the new label and with
the results from applying the function recursively to all subtrees.

\begin{figure}[here]
\centering
\begin{tikzpicture}
  \node (n1) [shape=circle,draw] at (-2,2)   {} ;
  \node (n2) [shape=circle,fill] at (2,2)    {} ;
  \node (n3) [shape=circle,draw] at (-2.5,1) {} ;
  \node (n4) [shape=circle,draw] at (-1.5,1) {} ;
  \node (n5) [shape=circle,fill] at ( 1.5,1) {} ;
  \node (n6) [shape=circle,draw] at ( 2.5,1) {} ;
  \node (n7) [shape=circle,draw] at ( -2.75,0) {} ;
  \node (n8) [shape=circle,draw] at ( -2.25,0) {} ;
  \node (n9) at ( -1.75,0) {5} ;
  \node (na) at ( -1.25,0) {6} ;
  \node (nb) [shape=circle,draw] at ( 1.75,0) {} ;
  \node (nc) [shape=circle,draw] at ( 1.25,0) {} ;
  \node (nd) at ( 2.75,0) {4} ;
  \node (ne) at ( 2.25,0) {2} ;
  \node (nf) at (-2.9,-1) {1} ;
  \node (ng) at (-2.6,-1) {3} ;
  \node (nh) at (-2.4,-1) {2} ;
  \node (ni) at (-2.1,-1) {4} ;
  \node (nj) at (1.6,-1) {1} ;
  \node (nk) at (1.9,-1) {3} ;
  \node (nl) at (1.1,-1) {5} ;
  \node (nm) at (1.4,-1) {6} ;
  \draw [->] (n1) -- (n3) ;
  \draw [->] (n1) -- (n4) ;
  \draw [->] (n2) -- (n5) ;
  \draw [->] (n2) -- (n6) ;
  \draw [->] (n3) -- (n7) ;
  \draw [->] (n3) -- (n8) ;
  \draw [->] (n4) -- (n9) ;
  \draw [->] (n4) -- (na) ;
  \draw [->] (n5) -- (nb) ;
  \draw [->] (n5) -- (nc) ;
  \draw [->] (n6) -- (nd) ;
  \draw [->] (n6) -- (ne) ;
  \draw [->] (n7) -- (nf) ;
  \draw [->] (n7) -- (ng) ;
  \draw [->] (n8) -- (nh) ;
  \draw [->] (n8) -- (ni) ;
  \draw [->] (nb) -- (nj) ;
  \draw [->] (nb) -- (nk) ;
  \draw [->] (nc) -- (nl) ;
  \draw [->] (nc) -- (nm) ;
\end{tikzpicture}
\caption{A tree monomial, and a decorated tree which is not a tree monomial} 
\label{fig:treeshape3}
\end{figure}
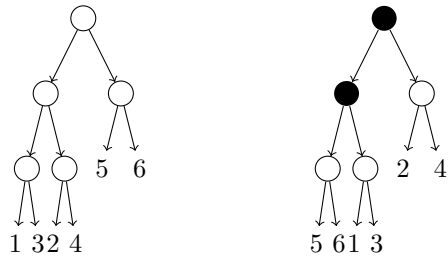

This structure faithfully reproduces planar rooted trees with leaf and
node labels. Using the ordering on the leaf labels, we can represent
any symmetric node-labeled tree this way. Using only the tree monomials 
from the free shuffle operad, finally, means placing restrictions on the permutations the
leaf labels are allowed to display. Specifically, at any vertex, the minimal leaves of all its subtrees need to occur in sorted
order in its list; this is exactly the choice of a planar representative from Section~\ref{sec:trees}. Hence, in Figure~\ref{fig:treeshape3}, the left tree is a valid tree monomial, whereas the right tree is not. The points where
the assumption fails are denoted by filled circles.

\subsection{Operad elements}
\label{sec:operad-elements}

Recall that an element of the free operad in the category of vector
spaces over a field is a formal linear combination of decorated
trees. To represent operad elements, thus, we need to be able to
represent formal linear combinations~--- sorted according to the
appropriate monomial ordering.

Representing this computationally in Haskell is a three-step
process. First, we represent monomial orderings. Then, we represent
trees equipped with a monomial ordering. Finally, we use the
\lstinline!Data.Map! Haskell standard library implementation to
represent a partially defined function taking decorated trees to
coefficient values.

These last functions become equivalent with formal linear combinations
once we define an arithmetic on them: 

\begin{align*}
  (f+g)(\alpha) &= f(\alpha) + g(\alpha) \\
  (c\cdot f)(\alpha) &= c\cdot f(\alpha)
\end{align*}

This equivalence is given by, for a function $f$, forming the formal
linear combination $\sum f(\alpha) \alpha$. The reverse is given by taking a
formal linear combination $\sum c_\alpha \alpha$ and forming the function $f: \alpha
\mapsto c_\alpha$. 

Monomial orderings are represented as \emph{empty types}~---
constructors without values other than having distinct
constructors. These types, then, are made to implement \emph{type
  classes}~--- the Haskell way to do polymorphism. Implementing a type
class means that the type class defines specific functions, and the
type class implementation defines the implementation of these
functions, in a separate manner for each type that implements the type
class. We pair trees with orderings using the record type facility~---
a tree with an ordering is a tree paired with an ordering. A number of
easy conversion functions between ordered and non-ordered trees make
interfacing with this layering easier.

As for the partial function definition, the datatype
\lstinline!Data.Map! works for finite such definitions by way of a
lookup table: an entity of type \lstinline!Data.Map! is a search tree
that can be queried for the value associated to a particular tree, and
that works, internally, by maintaining a balanced binary search
tree. In particular, this makes the use of \lstinline!Data.Map! very
dependent on an efficient implementation of the monomial ordering
methods, and one early adjustement we decided on was to overlay a
thin, encapsulating module around \lstinline!Data.Map! that would
cache the relevant information needed to perform the most common
monomial orderings.

This last point is worth elaborating on. We found in early
implementations that storing decorated trees in a binary search tree,
and having monomial orderings depending on a significant number of
tree traversals in order to construct the ordering invariants, lead to
an extraordinary amount of tree traversals. In our first working
implementation, we found that over 60\% of the computation time was
spent traversing trees for comparisons triggered by the use of
\lstinline!Data.Map! for storage. Every operation on such operad
elements would incur many tree comparisons, each of which would
incur several tree traversals. To deal with this, we wrote a wrapper
around the storage type that would perform the tree traversals for the
orderings described in Section \ref{sec:monomial-orderings} once at
the creation of an operad element, and store this with the
tree. Subsequent interactions with this particular element would use
this cached value for all comparisons~--- until the point a new
element was constructed, as a modification of the previous one, at
which point the comparison values would be recomputed. Using the
wrapper, the proportion spent on tree comparisons and the related
building block functions dropped to less than 5\%.

\subsection{Trees with holes and tagged nodes}
\label{sec:trees-with-holes}

Recall that for two tree monomials $\alpha$ and $\beta$ such that $\alpha$ is divisible by~$\beta$,
it is possible to define the function $m_{\alpha,\beta}$ that reconstructs the surroundings of~$\beta$ 
in~$\alpha$; this function is applicable to any other operad element of the correct arity.

An algorithm for finding divisors of trees, as the Gr\"obner basis
algorithms makes heavy use of, would need an efficient way to
represent the data needed for such a reconstruction. We decided to do
this by representing holes punched in trees by corollas with a
specific marking.

Similar to this idea is the data representation we found to be
efficient to represent a small common multiple of two trees in such a
manner that the divisor data for both trees is easily reconstructed:
such a small common multiple will have one of the trees dividing it at
the root, and the other somewhere in the tree. We found a natural way,
in Haskell, using union data types, to represent a tree with a single
vertex marked.

\subsubsection{An aside on data types and labels}
\label{sec:an-aside-data}

The union data type construction in Haskell has a standard library
implementation, with quite a bit of predefined functionality:
\lstinline!Either a b!. This defines for us constructors
\lstinline!Left! and \lstinline!Right! carrying values of types
\lstinline!a! and \lstinline!b! respectively.

One special case of the \lstinline!Either! type is when one of the two
types is empty. This case has been given a name of its own:
\lstinline!Maybe a! and comes with new constructors~---
\lstinline!Just! taking the place of \lstinline!Left! and
\lstinline!Nothing! taking the place of \lstinline!Right!.

These type constructions turn out to be exactly what we need to
signify marked nodes and removed nodes.

\subsubsection{Trees with marked nodes}
\label{sec:trees-with-marked}

We generate, in order to mark some of the nodes of our tree, a new
tree from the old one with changed node labels. Instead of labelling
our tree with some type \lstinline!a!, we now label them with
\lstinline!Either a a!. This has the effect of increasing the amount
of information carried by each node~--- in addition to the original
node information it now also carries a binary choice for each node: is
it a \lstinline!Left! or a \lstinline!Right! instance of the label
type?

Hence, we can in our code for generating small common multiples return
a tree labeled in \lstinline!Either a a!, and making sure it only
contains one single node labeled \lstinline!Right!~--- namely the point
of attachment for the second tree. We know that one of the two trees
has to be rooted at the root of the small common multiple~---
otherwise it would not cover all vertices.

The algorithm we use to find small common multiples, elaborated on in
Section \ref{sec:finding-small-common}, has the following basic
structure. In order to find small common multiples of $\alpha_1$ and $\alpha_2$ we
go through the steps:
\begin{enumerate}
\item To find small common multiples of $\alpha_1$ and $\alpha_2$ sharing the common
  root, traverse both trees, checking compatibility at each step
  and whenever one tree yields a leaf, attach the remaining subtree of
  the other tree. Tag the root of the returned small common multiple
  as \lstinline!Right label!.
\item Recurse through vertices~$v$ of $\alpha_2$, applying the previous step to find
small common multiples of~$\alpha_1$ and the subtree~$\alpha_2^v$ of~$\alpha_2$ rooted at~$v$.
For each such common multiple~$\gamma$, form a new tree by taking $\alpha_2$ and replacing $\alpha_2^v$ 
with~$\gamma$. 
\end{enumerate}

As a result, the only point where a node is tagged with the
\lstinline!Right! is when a rooted common multiple is found, and the
recursion ensures that the rest of the tree is rebuilt so that
$\alpha_2$ is embedded with a shared root with the common
multiple.

And in order to find all small common multiples, we need to perform
this algorithm once again with the trees interchanged, so that we may
find small common multiples with $\alpha_1$ embedded at the root.

\subsubsection{Trees with holes}
\label{sec:trees-with-holes-2}

As for the divisor reconstruction, we have some embedding of the tree monomial
$\beta$ into the tree $\alpha$, and we want to retain the information of the
entire tree excepting the part that corresponds to $\beta$. 

The way we do this is to collapse the embedded copy of $\beta$ into a
single corolla of the correct arity, and keeping the rest of the tree
intact. This corolla, then, is marked~--- forgetting any original
corolla type markings~--- to signify that it forms an embedding point.

This marking, in turn, is achieved by changing the type of all labels
from the type \lstinline!a! to the type \lstinline!Maybe a!. That way,
the part of the tree that needs to stay intact is marked with
\lstinline!Just label! for what previously was marked with
\lstinline!label!, and the corolla holding the position of the hole is
marked with \lstinline!Nothing!, distinguishing it from the other
nodes.

\begin{figure}[here]
\centering
\begin{tikzpicture}[thin]
  \begin{scope}[xshift=-3cm]
    \node [draw,shape=circle] (rt) at (0,0) {} ;
    \node [draw,shape=circle,blue,very thick] (l) at (-0.7,-1) {} ;
    \node [draw,shape=circle] (r) at (0.7,-1) {} ;
    \node [draw,shape=circle,blue,very thick] (ll) at (-1.2,-2) {} ; 
    \node [blue,very thick] (lm) at (-0.7,-2) {3} ; 
    \node [draw,shape=circle,blue,very thick] (lr) at (-0.2,-2) {} ;
    \node (rl) at (0.2,-2) {6} ;
    \node [draw,shape=circle] (rr) at (1.2,-2) {} ;
    \node (lll) at (-1.5,-3) {1} ;
    \node (llr) at (-0.9,-3) {2} ;
    \node [blue,very thick] (lrl) at (-0.5,-3) {4} ;
    \node [blue,very thick] (lrr) at (0.1,-3) {5} ;
    \node (rrl) at (0.8,-3) {7} ;
    \node (rrm) at (1.2,-3) {8} ;
    \node (rrr) at (1.6,-3) {9} ;
    
    \draw [->] (rt) -- (l) ;
    \draw [->] (rt) -- (r) ;
    \draw [->,blue,very thick] (l) -- (ll) ;
    \draw [->,blue,very thick] (l) -- (lm) ;
    \draw [->,blue,very thick] (l) -- (lr) ;
    \draw [->] (r) -- (rl) ;
    \draw [->] (r) -- (rr) ;
    \draw [->] (ll) -- (lll) ;
    \draw [->] (ll) -- (llr) ;
    \draw [->,blue,very thick] (lr) -- (lrl) ;
    \draw [->,blue,very thick] (lr) -- (lrr) ;
    \draw [->] (rr) -- (rrl) ;
    \draw [->] (rr) -- (rrm) ;
    \draw [->] (rr) -- (rrr) ;
  \end{scope}
  \begin{scope}[xshift=3cm]
    \node [draw,shape=circle] (rt) at (0,0) {} ;
    \node [draw,shape=rectangle,blue,very thick,scale=2] (l) at (-0.7,-1) {} ;
    \node [draw,shape=circle] (r) at (0.7,-1) {} ;
    \node [draw,shape=circle] (ll) at (-1.5,-2) {} ; 
    \node (lm) at (-1.0,-2) {3} ; 
    \node (rl) at (0.3,-2) {6} ;
    \node [draw,shape=circle] (rr) at (1.2,-2) {} ;
    \node (lll) at (-1.7,-3) {1} ;
    \node (llr) at (-1.3,-3) {2} ;
    \node (lrl) at (-0.6,-2) {4} ;
    \node (lrr) at (-0.2,-2) {5} ;
    \node (rrl) at (0.8,-3) {7} ;
    \node (rrm) at (1.2,-3) {8} ;
    \node (rrr) at (1.6,-3) {9} ;
    
    \draw [->] (rt) -- (l) ;
    \draw [->] (rt) -- (r) ;
    \draw [->] (l) -- (ll) ;
    \draw [->] (l) -- (lm) ;
    \draw [->] (l) -- (lrl) ;
    \draw [->] (l) -- (lrr) ;
    \draw [->] (r) -- (rl) ;
    \draw [->] (r) -- (rr) ;
    \draw [->] (ll) -- (lll) ;
    \draw [->] (ll) -- (llr) ;
    \draw [->] (rr) -- (rrl) ;
    \draw [->] (rr) -- (rrm) ;
    \draw [->] (rr) -- (rrr) ;
  \end{scope}
\end{tikzpicture}
\caption{Taking away a subtree results in a tree with a hole} 
\label{fig:treeshape4}
\end{figure}
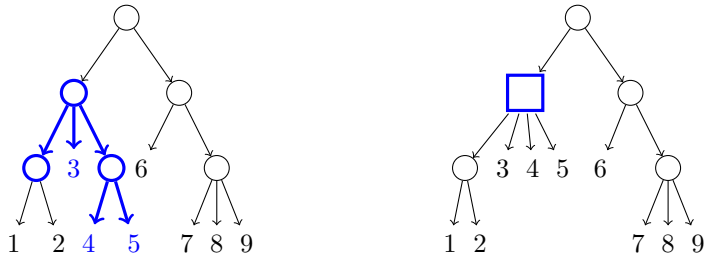

For this reason, we introduce the type alias \lstinline!Embedding a!,
defined to be of the type \lstinline!DecoratedTree (Maybe a)!. Hence,
the division algorithm described in Section \ref{sec:divis-reconstr}
will take the two trees $\alpha$ and $\beta$ as parameters, and return an
embedding of the shape of $\alpha$ with a subtree isomorphic to $\beta$ taken
out. See Figure~\ref{fig:treeshape4} for an example.

The reconstruction algorithm, on the other hand, takes a shape representing some
embedding of $\beta$ in $\alpha$, and a new tree monomial $\gamma$ of the same arity as $\beta$, and
returns the corresponding tree $m_{\alpha,\beta}(\gamma)$.

\subsection{Permutations}
\label{sec:permutations}

Since the most common use for permutations in this project is to label
leaves of trees, and to reorder subtrees for composition, we have
decided to store our permutations as lists of images. This choice is
reinforced by the lack of need for compositions and decompositions 
of permutations.

This representation yields a simple method to reorder a list of
objects in the order specified by a permutation~--- an operation we have
reason to perform often in the code, for instance in order to decorate
leaves of a labelled tree according to their integer decorations: we
pair off the elements we want to reorder with the image list. Then we
sort the pairs, giving priority to the comparison of the image
indices. Stripping off the indices, finally, gives us the reordered
list of elements.

This is a code idiom we have used at several points in the code
base.

\section{Algorithms}
\label{sec:algorithms}

With the data structures we use settled in Section
\ref{sec:intern-repr}, we now turn to the algorithms that implement
the core components of the Buchberger algorithm. Thus, in Section
\ref{sec:divis-reconstr}, we meet the division algorithm, creating the
black hole trees, and the reconstruction algorithm, re-inserting a tree
in the black hole. In Section \ref{sec:finding-small-common}, we adapt
the idea for finding block permutations to give us an efficient
algorithm for finding small common multiples. Finally, in Section
\ref{sec:monomial-orderings} we discuss the family of monomial
orderings that the software package implements.

\subsection{Divisibility and reconstructions}
\label{sec:divis-reconstr}

Finding all embeddings (as a divisor) of $\beta$ into $\alpha$ can be easily reduced to finding out whether or not $\beta$ is embedded into $\alpha$ in 
such a way that they share the common root. If we know how to solve this problem, we should just solve it for all subtrees $\alpha'$ of $\alpha$ rooted at 
various vertices. To solve this problem, it suffices to recurse down the tree checking that the same holds for all subtrees, and then check that 
the total leaf orderings match. At this stage, to check that the orderings match, one can look at the planar orders of leaves and compare them; this 
appears to be the best way to do it for our recursive algorithm.

Specifically, if we try to find an embedding of $\beta$ into $\alpha$ sharing the common root, we first verify that the root vertices share the label and arity. If this is the case, we can pair up the subtrees $\beta_i$ of $\beta$ and the subtrees of $\alpha_i$ of 
$\alpha$, and then find rooted embeddings of each $\beta_i$ in each $\alpha_i$. If all these succeed, we expect to get as a result from each a tree 
$m_{\alpha_i,\beta_i}$ for each pair with a hole punched out at the root corresponding to a subtree looking like $\beta_i$.

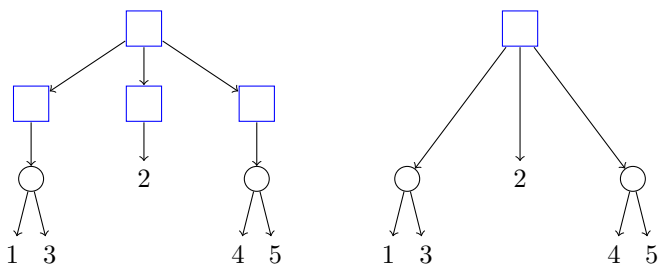
\begin{figure}[here]
\centering
\begin{tikzpicture}
  \tikzstyle{node} = [draw,circle]
  \tikzstyle{empty} = [draw,rectangle,scale=2,blue]
  \begin{scope}[xshift=-2.5cm]
    \node [empty]   (rt) at (0,0)       {}  ;
    \node [empty]   (le) at (-1.5,-1)   {}  ;
    \node [empty]   (me) at (0,-1)   {}  ;
    \node [empty]   (re) at (1.5,-1)   {}  ;
    \node [node]  (l)  at (-1.5,-2)   {}  ;
    \node           (m)  at (0,-2)      {2} ;
    \node [node]  (r)  at (1.5,-2)    {}  ;
    \node           (ll) at (-1.75,-3)  {1} ;
    \node           (lr) at (-1.25,-3)  {3} ;
    \node           (rl) at (1.25,-3)   {4} ;
    \node           (rr) at (1.75,-3)   {5} ;
    \draw [->] (rt) -- (le) ;
    \draw [->] (rt) -- (me) ;
    \draw [->] (rt) -- (re) ;
    \draw [->] (le) -- (l) ;
    \draw [->] (me) -- (m) ;
    \draw [->] (re) -- (r) ;
    \draw [->] (l) -- (ll) ;
    \draw [->] (l) -- (lr) ;
    \draw [->] (r) -- (rl) ;
    \draw [->] (r) -- (rr) ;
  \end{scope}
  \begin{scope}[xshift=2.5cm]
    \node [empty]   (rt) at (0,0)       {}  ;
    \node [node]  (l)  at (-1.5,-2)   {}  ;
    \node           (m)  at (0,-2)      {2} ;
    \node [node]  (r)  at (1.5,-2)    {}  ;
    \node           (ll) at (-1.75,-3)  {1} ;
    \node           (lr) at (-1.25,-3)  {3} ;
    \node           (rl) at (1.25,-3)   {4} ;
    \node           (rr) at (1.75,-3)   {5} ;
    \draw [->] (rt) -- (l) ;
    \draw [->] (rt) -- (m) ;
    \draw [->] (rt) -- (r) ;
    \draw [->] (l) -- (ll) ;
    \draw [->] (l) -- (lr) ;
    \draw [->] (r) -- (rl) ;
    \draw [->] (r) -- (rr) ;
  \end{scope}
\end{tikzpicture}
\caption{Merging subtrees with holes} 
\label{fig:treeshape5}
\end{figure}

At this point, we need to patch things up. The root nodes match, and the subtrees have already found embeddings. Checking the leaf orders, we then need to 
merge all the subtrees with holes into a tree with hole that gets returned up to earlier recursion levels. This is done by simply creating a new hole 
vertex, and then attaching all subtrees of the hole subtrees, as shown in Figure~\ref{fig:treeshape5}. 

Once we find an embedding of $\beta$, we store this embedding as a tree monomial obtained from~$\alpha$ by collapsing the occurrence of~$\beta$ into a 
single vertex. To reconstruct $\alpha$ from that, we insert $\beta$ in the ``hole'' in such a way that the leaves of~$\beta$ match the outputs of the ``hole'' 
(order-wise).

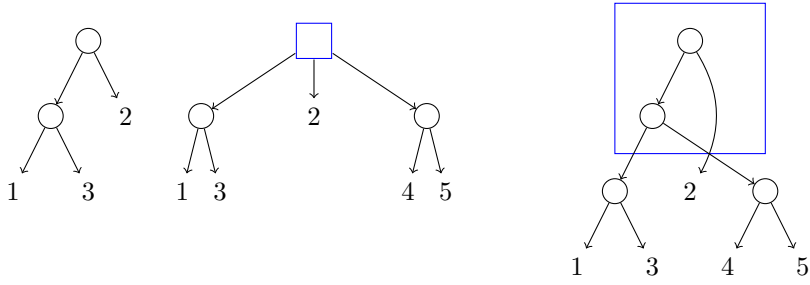
\begin{figure}[here]
\centering
\begin{tikzpicture}
  \tikzstyle{node} = [draw,circle]
  \tikzstyle{empty} = [draw,rectangle,scale=2,blue]
  \begin{scope}[xshift=-4cm]
    \node [node] (rt) at (0,0) {} ;
    \node [node] (l) at (-0.5,-1) {} ;
    \node (ll) at (-1,-2) {1} ;
    \node (lr) at (0,-2) {3} ;
    \node (r) at (0.5,-1) {2} ;
    \draw [->] (rt) -- (l) ;
    \draw [->] (rt) -- (r) ;
    \draw [->] (l) -- (ll) ;
    \draw [->] (l) -- (lr) ;
  \end{scope}
  \begin{scope}[xshift=-1cm]
    \node [empty]   (rt) at (0,0)       {}  ;
    \node [node]  (l)  at (-1.5,-1)   {}  ;
    \node           (m)  at (0,-1)      {2} ;
    \node [node]  (r)  at (1.5,-1)    {}  ;
    \node           (ll) at (-1.75,-2)  {1} ;
    \node           (lr) at (-1.25,-2)  {3} ;
    \node           (rl) at (1.25,-2)   {4} ;
    \node           (rr) at (1.75,-2)   {5} ;
    \draw [->] (rt) -- (l) ;
    \draw [->] (rt) -- (m) ;
    \draw [->] (rt) -- (r) ;
    \draw [->] (l) -- (ll) ;
    \draw [->] (l) -- (lr) ;
    \draw [->] (r) -- (rl) ;
    \draw [->] (r) -- (rr) ;
  \end{scope}
  \begin{scope}[xshift=4cm]
    \draw [blue] (-1,0.5) rectangle (1,-1.5);
    \node [node] (rt) at (0,0) {} ;
    \node [node] (l) at (-0.5,-1) {} ;
    \node [node] (ll) at (-1,-2) {} ;
    \node (lr) at (0,-2) {2} ;
    \node [node] (r) at (1,-2) {} ;
    \node (lll) at (-1.5,-3) {1} ;
    \node (llr) at (-0.5,-3) {3} ;
    \node (rl) at (0.5,-3) {4} ;
    \node (rr) at (1.5,-3) {5} ;

    \draw [->] (rt) -- (l) ; 
    \draw [->] (rt) to [bend left] (lr) ; 
    \draw [->] (l) -- (ll) ; 
    \draw [->] (l) to (r) ; 
    \draw [->] (ll) -- (lll) ;
    \draw [->] (ll) -- (llr) ;
    \draw [->] (r) -- (rl) ;
    \draw [->] (r) -- (rr) ;
  \end{scope}
\end{tikzpicture}
\caption{Inserting a tree in a hole} 
\label{fig:treeshape6}
\end{figure}

Inserting $\beta$ into the hole, specifically, means that we replace the leaves of $\beta$ with the subtrees of the hole in the tree with the hole, in the 
order specified by the labels of the leaves of $\beta$, and then replace the hole and all its subtrees in the full tree with a hole by this extended 
$\beta$, as shown in Figure~\ref{fig:treeshape6}.

\subsection{Finding small common multiples}
\label{sec:finding-small-common}

An algorithm that lists all small common multiples of two given trees $\alpha_1$ and~$\alpha_2$ consists of several steps. If we forget all leaf labels of a tree 
monomial, we end up with a planar tree with labelled vertices. We call such tree a nonsymmetric tree monomial; such trees form a basis in the 
free nonsymmetric operad generated by the same nonsymmetric collection as our free shuffle operad. Describing all small common multiples is naturally split 
into two steps: forgetting about leaf labels and finding a nonsymmetric small common multiple, and then acquiring all possible leaf labellings of the 
resulting trees.

The first step is more or less trivial: small common multiples of two nonsymmetric tree monomials are superpositions of the trees for which all labels of vertices agree with each other. Thus, to list all such small common multiples, we should go through all ways to identify the root vertex of one of 
the trees with a vertex of another tree, and check that there are no contradictions between the successors of these two vertices. This can be 
easily done recursively.

The second step is a bit more tricky, but still requires a straightforward recursive algorithm. To recover all admissible leaf labellings giving a common multiple~$\alpha$, we have to solve the problem of finding all possible linear orders on a poset of a special type. The elements of that poset are leaves 
of the nonsymmetric tree monomial, and we say that $a<b$ if there exist two vertices $u$ and $v$ of the nonsymmetric tree monomial such that

\begin{itemize}
 \item $a$ is the smallest leaf reachable from~$u$ and $b$ is the smallest leaf reachable from~$v$;
 \item $u$ and $v$ are leaves of the occurence of~$\alpha_i$ in~$\alpha$ (where $i$ is either~$1$ or~$2$), and $u<v$ in the ordered leaf set of~$\alpha_i$.
\end{itemize}

A labelling of the leaves of~$\alpha$ makes it a small common multiple if and only if it extends the above ordering to a linear ordering. 
We shall recover all such labellings recursively. Our poset essentially consists of two intertwined and intersecting linear orders, and the maximal element 
of the labelling set should label the maximal elements of one of the two orders (under the additional condition that it cannot occur as a non-maximal 
element in the other linear order). For each such option, we are left with a similar problem where the size of the labelling set is one less, so we can use 
recursion.

\subsection{Monomial orderings}
\label{sec:monomial-orderings}

Let us describe some admissible orderings. As one can see, each definition will be either immediate to implement because of the storage types we use or 
straightforward recursive.

Let $\alpha$ be a tree monomial with $n$ inputs in the free operad~$\calF_M$. We associate to $\alpha$ a sequence $(a_1,a_2, \ldots,a_n)$ of $n$ words in 
the alphabet $M$, and a permutation $g\in S_n$ as follows. For each leaf~$i$ of~$\alpha$, there exists a unique path from the root to~$i$. The word $a_i$ 
is the word composed, from left to right, of the labels of the vertices of this path, starting from the root vertex. The permutation $g$ lists the labels 
of leaves in the order determined by the planar structure (from left to right).

Now, to compare two tree monomials we always compare their arities first. If the arities are equal, there are several different options of how to proceed. 
Sequences of words can be compared lexicographically using either the degree-lexicographic ordering of words, or the reverse degree-lexicographic ordering 
(either the longer word is greater, or vice versa; for words of the same length the comparison is lexicographic). Permutations can be compared in the 
lexicographic or reverse lexicographic order. Also, the result depends on what we compare first, the permutations or the sequences of words. This gives 
rise to eight candidates for an ordering; we name these candidates \textbf{PathPerm}, \textbf{RPathPerm}, \textbf{PathRPerm}, \textbf{RPathRPerm}, 
\textbf{PermPath} etc. (the names are self-explanatory).

\begin{proposition}[\cite{dotsenko_freeness}]
All the above orderings are admissible.
\end{proposition}

In fact, to compare words one may use any admissible ordering of the monomial basis of the free algebra, for example, the lexicographic ordering, or
the reverse lexicographic one: the resulting ordering of tree monomials will be admissible as well.

\section*{Appendix: Haskell constructions used}
\label{sec:hask-constr-used}

\begin{description}
\item[Bool] The boolean truth values type. Has values
  \lstinline!True! and \lstinline!False!.
\item[Int] The bounded integer type. Has values, on a 32
  bit machine, from the interval $[-2147483648,2147483647]$.
\item[Char] The single character type.
\item[{[a]}] The type of lists of elements of type \textbf{a}.
\item[a $\rightarrow$ b] The type of a function from a type \textbf{a} to a
  type \textbf{b}.
\item[data] The declaration of a new data type.
\item[Ord] The type class that defines the ordering functions
  \verb!<!, \verb!>!, \verb!<=!, \verb!>=!, and \verb!compare!.
\item[Eq] The type class that defines the equality testing function
  (\verb!==!).
\item[Show] The type class that defines the serialization function
  \lstinline!show!.
\item[Fractional] The type class that defines a type to implement a
  field.
\item[(::)] The syntax element indicating a type declaration.
\item[($\Rightarrow$)] The syntax element delimiting type assumptions from the
  type declaration.
\item[(!)] When occurring in a type declaration, forces strictness in
  the corresponding part.
\item[($|$)] When occurring in a type declaration, delimiting the union
  type components. Hence, a type declared as 
  \lstinline!data T = A Int | B Char!
  is either an \lstinline!Int! with the constructor \lstinline!A!, or
  a \lstinline!Char! with the constructor \lstinline!B!.
\item[TreeOrdering] A type class created by our code carrying
  information about the chosen monomial order.
\item[OperadElement] The type created by our code representing,
  internally, a linear combination of tree monomials with associated
  monomial orderings.
\item[$(+)$] The addition function.
\item[deriving] Used in a \textbf{data} declaration. It will
  automatically generate implementations of the type classes listed.
\item[map] A higher order function that applies another function to
  every element in a list.
\end{description}

\bibliographystyle{plain}
\bibliography{igb}

\end{document}